\documentclass[submission,copyright,creativecommons]{eptcs}
\providecommand{\event}{ICLP 2026} 

\usepackage[utf8]{inputenc}

\usepackage{todonotes}
\usepackage{fourier-orns}

\usepackage{amsmath}
\usepackage{algorithmicx}
\usepackage[ruled]{algorithm}
\usepackage[noend]{algpseudocode}
\usepackage{pmboxdraw} 
\usepackage[misc]{ifsym}
\usepackage{subcaption}
\usepackage{enumitem}
\usepackage{bsymb}
\usepackage{proof} 
\usepackage{wrapfig}
\usepackage{makecell}

\DeclareUnicodeCharacter{2119}{$\mathbb{P}$}
\DeclareUnicodeCharacter{2124}{$\mathbb{Z}$}
\DeclareUnicodeCharacter{2194}{$\leftrightarrow$}
\DeclareUnicodeCharacter{21F8}{$\pfun{}$}
\DeclareUnicodeCharacter{2208}{$\in$}
\DeclareUnicodeCharacter{2227}{$\land$}
\DeclareUnicodeCharacter{2286}{$\subseteq$}

\newcommand{\prob}{\textsc{ProB}}
\newcommand{\probtwoui}{\textsc{ProB2-UI}}
\newcommand{\atelierb}{\textsc{Atelier-B}}
\newcommand{\rodin}{\textsc{Rodin}}

\newcommand{\visb}{\textsc{VisB}}
\newcommand{\simb}{\textsc{SimB}}

\newcommand{\ignore}[1]{}
\newcommand{\condtext}[2]{#1}

\usepackage{hyperref}
\newcommand{\urlhref}[2]{%
	\href{#1}{\textcolor{blue}{#2}}%
}
\usepackage[noabbrev]{cleveref}
\usepackage{color}

\usepackage{xcolor}
\definecolor{orcidlogocol}{HTML}{A6CE39}
\definecolor{prob2uicolor}{HTML}{037875}
\usetikzlibrary{svg.path,calc}
\tikzset{
	orcidlogo/.pic={
		\fill[orcidlogocol] svg{M256,128c0,70.7-57.3,128-128,128C57.3,256,0,198.7,0,128C0,57.3,57.3,0,128,0C198.7,0,256,57.3,256,128z};
		\fill[white] svg{M86.3,186.2H70.9V79.1h15.4v48.4V186.2z}
		svg{M108.9,79.1h41.6c39.6,0,57,28.3,57,53.6c0,27.5-21.5,53.6-56.8,53.6h-41.8V79.1z M124.3,172.4h24.5c34.9,0,42.9-26.5,42.9-39.7c0-21.5-13.7-39.7-43.7-39.7h-23.7V172.4z}
		svg{M88.7,56.8c0,5.5-4.5,10.1-10.1,10.1c-5.6,0-10.1-4.6-10.1-10.1c0-5.6,4.5-10.1,10.1-10.1C84.2,46.7,88.7,51.3,88.7,56.8z};
	}
}
\newcommand{\orcid}[1]{%
	\resizebox{8px}{8px}{
		\href{https://orcid.org/#1}{\tikz[yscale=-1,transform shape]{\pic{orcidlogo}}}}%
}
\usepackage[final,frozencache,cachedir=minted-cache]{minted2}

\DeclareCaptionType{mintlisting}[Listing][List of Listings]
\counterwithout{mintlisting}{section}
\crefname{mintlisting}{Listing}{Listings}
\Crefname{mintlisting}{Listing}{Listings}

\title{Encoding Event-B Proof Rules in Prolog:\\An Interactive Sequent Prover for \prob{}}
\author{Katharina Engels \orcid{0009-0001-3595-3683} \qquad \quad Jan Gruteser \orcid{0009-0006-4228-404X} \qquad\quad Michael Leuschel \orcid{0000-0002-4595-1518}
	\institute{Faculty of Mathematics and Natural Science, Institute of Computer Science,\\
		Heinrich Heine University Düsseldorf, Universit\"{a}tsstr. 1, D-40225 D\"{u}sseldorf}
	\email{\quad \{katharina.engels,jan.gruteser,michael.leuschel\}@hhu.de}
}
\def\titlerunning{Encoding Event-B Proof Rules in Prolog: An Interactive Sequent Prover for \prob{}}
\def\authorrunning{K. Engels, J. Gruteser, and M. Leuschel}

\begin{document}
\maketitle

\begin{abstract}
Event-B is a formal method rooted in predicate logic and set theory.
We encoded 
over 600 proof rules in Prolog, enabling a systematic, comprehensible proof analysis and construction.
By integrating the proof rules into the Prolog-based validation tool \prob{}, we 
 obtain an interactive proof system with proof tree visualisation.
This has advantages in teaching, giving students direct control over the selection of proof rules.
Our tool can import proof obligations from the \rodin{} platform and provides multiple exports:
 a trace file for proof replay in \prob{}, an interactive HTML document for tool-independent exploration of the proof tree, 
 and an export 
  back to \rodin{}, allowing the \prob{} prover to be used as second chain.
Compared to the previous implementation of the proof rules in Java, the encoding in Prolog is more compact,
 maintainable and extensible.
While a preliminary iterative deepening prover with simple heuristics is already available and useful for finding short proofs, we aim to obtain fast automatic provers in the future.
\end{abstract}


\section{Introduction and Motivation}


In this work, we have developed a new interactive and automated theorem prover in Prolog for the Event-B formal method.
The Event-B language introduced by Abrial~\cite{bmethod,Abrial10} is founded on predicate logic and set theory and builds on invariants and refinement to develop systems that are correct by construction, where proving is a crucial part to mathematically verify the system's correctness.
The logic of Event-B is underpinned by over 600 proof rules in sequent calculus style~\cite{gentzen1935untersuchungen}.
These proof rules are embedded in the \rodin{} platform~\cite{rodin},
 which also caters for several industrial and academic plugins to automatically conduct proofs.
The \rodin{} platform contains the proof obligation generator, which generates proof obligations (POs) for an Event-B model.
Discharging all POs ensures consistency and correctness of the model.

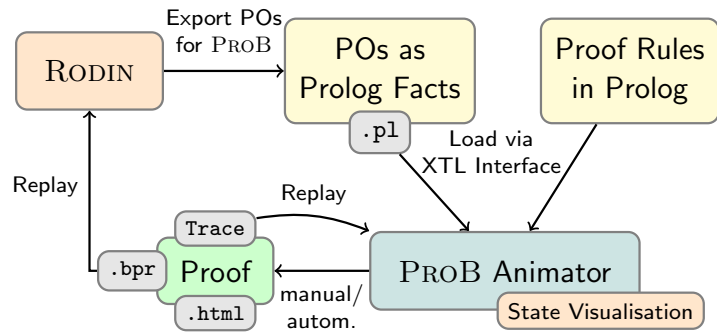
\begin{wrapfigure}[11]{r}{0.6\linewidth}
	\vspace{-0.8cm}
	\centering
	\resizebox{\linewidth}{!}{%
		\begin{tikzpicture}[auto, every loop/.style={},
			thick,
			node/.style={rectangle,font=\sffamily\normalsize},
			edge/.style={font=\sffamily\scriptsize}
			]<
			
			\node[node, rounded corners, draw=black!50, fill=orange!20, inner sep=10pt] (rodin) at (-0.075,0) {\rodin};
			
			\node[node, rounded corners, draw=black!50, fill=yellow!20, right=1.5cm of rodin, inner sep=5pt] (pofile) {\makecell{POs as\\Prolog Facts}};
			\node[node, rounded corners, draw=black!50, fill=gray!20] (pl) at ($(pofile.south) - (0,0.1)$) {\footnotesize\texttt{.pl}};
			\node[node, rounded corners, draw=black!50, fill=gray!20, right=0.8cm of pofile, inner sep=5pt, fill=yellow!20] (rules) {\makecell{Proof Rules\\in Prolog}};
			
			\node[node, rounded corners, draw=black!50, fill=green!20, inner sep=8pt] (proof) at (1.5,-2.5) {Proof};
			\node[node, rounded corners, draw=black!50, fill=gray!20] (bpr) at ($(proof.west) - (0.3,0)$) {\scriptsize\texttt{.bpr}};
			\node[node, rounded corners, draw=black!50, fill=gray!20] (trace) at ($(proof.north) + (0,0.1)$) {\scriptsize\texttt{Trace}};
			\node[node, rounded corners, draw=black!50, fill=gray!20] (html) at ($(proof.south) - (0,0.1)$) {\scriptsize\texttt{.html}};
			
			\node[node, rounded corners, draw=black!50, fill=prob2uicolor!20, inner sep=10pt] (animator) at ($(rules |- proof)!0.5!(pofile |- proof)$) {\prob{} Animator};
			
			\node[node, rounded corners, draw=black!50, fill=orange!20] (simb) at ($(animator.south east) + (-0.5,0)$) {\scriptsize State Visualisation};
			
			\draw[edge, ->] (rodin) -- node[above] {\makecell{Export POs\\for \prob}} (pofile);
			\draw[edge, ->] (pl) -- node[above right, xshift=-3mm] {\makecell{Load via\\XTL Interface}} (animator);
			\draw[edge, ->] (rules) -- (animator);
			\draw[edge, ->] (animator) -- node[below] {\makecell{manual/\\autom.}} (proof);
			\draw[edge, ->] (trace) to[bend left=15] node[above] {Replay} (animator.north west);
			\draw[edge, ->] (bpr) -| node[left, yshift=30pt] {Replay} (rodin);
		\end{tikzpicture}
	}
	\caption{Architecture of \prob{}'s Sequent Prover}
	\label{fig:toolchain}
\end{wrapfigure}

The initial motivation of this work came from teaching.
While the Event-B proof rules are discussed in detail during lectures and are presented in the literature~\cite{Abrial10}, \rodin{} often applies several proof steps simultaneously. 
The plugins of \rodin{} also conduct an entire proof in a single step.
In both cases, the user or student cannot easily check, understand or reproduce a proof.
Hence, the initial intention was to develop an interactive prover for Event-B, enabling students to apply or replay proof rules step by step and gain experience through experimentation.
Our implementation, however, has a variety of other benefits for advanced users:
  a second independent toolchain to double-check proofs,
  new visualisation and analysis features,
  and a new automatic prover producing proofs that are more robust to model changes and easier to reproduce upon changes of the proof itself.

In this article, we show how the mathematical definitions of Event-B can be turned into executable Prolog rules that are still close to the original definition and can be used for proving.
This includes almost all of \rodin{}'s inference and rewrite rules\footnote{%
	\urlhref{https://wiki.event-b.org/index.php/Inference_Rules}{https://wiki.event-b.org/index.php/Inference\_Rules}/%
	\urlhref{https://wiki.event-b.org/index.php/All_Rewrite_Rules}{All\_Rewrite\_Rules}
}, but also a few additional rules proposed by Abrial~\cite{Abrial10}.
Within the \rodin{} platform, the rules are implemented in Java.
By re-implementing those proof rules in Prolog, we have obtained a more compact code base.
The encoding in Prolog is easier to maintain, extend for new proof rules and can be used for more purposes than the original Java code.
In particular, we can use Prolog's search capabilities for automatically generating proofs.

With the \prob{} validation tool \cite{DBLP:journals/sttt/LeuschelB08,leuschel2023prob} written in Prolog~(cf. \Cref{fig:toolchain}),
  we can bring the proof rules to life and turn
  them into a labelled transition system, where the sequents are the states and each application of a proof rule is a transition.
This allows students and regular users to choose which proof rule is applied at each proof step 
but also enables interactive HTML visualisations of finished proofs.
By generating \rodin{} proof files (BPR) in XML format, we close the loop to the \rodin{} prover, enabling the \prob{} sequent prover to be used as a second chain.

In this article, we present the following core contributions:
\begin{itemize}
    \item a Prolog implementation of Event-B proof inference and rewrite rules,
    \item a comparison of the existing Java code in the \rodin{} tool with our Prolog code regarding the code size and the effort required to implement a new rule,  
    \item the integration of the proof rules into the validation tool \prob{}, featuring
    \begin{itemize}
        \item an interactive sequent prover using the \prob{} animator, along with an interactive visualisation showing the current proof sequent,
        \item an export of a proof tree visualisation as interactive HTML document using \prob{}'s interface to Graphviz for detailed analysis and teaching,
        \item an export of a proof as BPR file for replay 
        in the \rodin{} platform,
    \end{itemize}
    \item a first version of an automated prover for later integration in \rodin{}, and
    \item a demonstration of the usefulness of the tooling for teaching proofs and logics in the context of formal modelling.
\end{itemize}


\section{Event-B Proofs}\label{sec:background}


In Event-B, proof obligations (POs) must be proven to verify the correctness of a model. 
Proof obligations arise, for example, from invariants, refinement, user-specified theorems or well-definedness conditions such as potential divisions by zero.\footnote{In the case of a PO for invariant preservation for some event, the hypotheses include defined axioms, 
 theorems and the invariants applied to the state variables before the event. 
 }

Depending on the type of a PO, a selection of hypotheses $H$ is available to prove the goal $G$.
Formally, a PO can be represented by a sequent $H \vdash G$.
This sequent expresses that the goal $G$ logically follows from the set of hypotheses $H$ using the proof rules of Event-B.



\label{sect-with-andr}
One can use a set of different inference rules to prove a sequent.
One of these rules is the \texttt{HYP} rule~($H, P \vdash P$), which states that a sequent is proven if the goal is contained in the hypotheses.
Two other inference rules are depicted below:
%
\[
\texttt{AND\_R}\;\; \;\frac{\textbf{H} \; \; \vdash \; \; \textbf{P} \qquad \textbf{H} \; \; \vdash \; \; \textbf{Q}}{\textbf{H} \; \; \vdash \; \; \textbf{P} \; \land \; \textbf{Q}}
\qquad
\frac{}{\textbf{H}, \; f \in E \tfun{} F\; \vdash \; f \in \textit{Type}_E \pfun \textit{Type}_F}\;\;\;\texttt{FUN\_GOAL}
\]
The left rule states that if the goal is a conjunction, 
the sequent can be proven by showing that each conjunct follows from the hypotheses individually. 
The right rule states that $f$ is a partial function ($\pfun$ symbol) if it is a total function $(\tfun$ symbol) for some domain $E$ and range $F$.

\begin{figure}[hb]
	\resizebox{\linewidth}{!}{%
		\begin{tikzpicture}[
			every node/.style={rectangle, align=left},
			arrow/.style={->, thick},
			smalllabel/.style = {font=\fontsize{7}{22.4}\selectfont}
			]
			\node[draw] (start) {radiation $\in$ field  $\tfun{}$ $\mathbb{N}$\\
				$\{$rover$\}$ $\subseteq$ field\\
				$\vdash$\\
				radiation $\in \mathbb{Z} \times \mathbb{Z} \pfun{} \mathbb{Z}$\\
				$\land$ rover $\in$ dom(radiation)};
			\node[right=1.2cm of start, yshift=1.3cm, draw] (and1) {
				radiation $\in$ field  $\tfun{}$ $\mathbb{N}$\\
				$\{$rover$\}$ $\subseteq$ field\\
				$\vdash$\\
				\textbf{radiation} $\pmb{\in \mathbb{Z} \times \mathbb{Z} \pfun{} \mathbb{Z}}$};
			\node[right=2.1cm of and1, draw] (success1) {Proven};
			\node[right=1.2cm of start, yshift=-1.3cm, draw] (and2) {
				radiation $\in$ field  $\tfun{}$ $\mathbb{N}$\\
				$\{$rover$\}$ $\subseteq$ field\\
				$\vdash$\\
				\textbf{rover} $\pmb{\in}$ \textbf{dom(radiation)}};
			\node[right=2.2cm of and2, draw] (deriv_dom) {
				radiation $\in$ field  $\tfun{}$ $\mathbb{N}$\\
				$\{$rover$\}$ $\subseteq$ field\\
				$\vdash$\\
				\textbf{rover} $\pmb{\in}$ \textbf{field}};
			\node[right=2.1cm of deriv_dom, draw] (simp_hyp) {
				radiation $\in$ field  $\tfun{}$ $\mathbb{N}$\\
				\textbf{rover} $\pmb{\in}$ \textbf{field}\\
				$\vdash$\\
				rover $\in$ field};
			\node[above=1.0cm of simp_hyp, draw] (success2) {Proven};
			
			\node[right=0.05cm of start, smalllabel] (and_label) {AND\_R};
			\node[above right=-0.8cm and 0.1cm of and1, smalllabel] (fun_label) {FUN\_GOAL};
			
			\draw[arrow] (start) -- (and1);
			\draw[arrow] (start) -- (and2);
			\draw[arrow] (and1) -- (success1);
			\draw[arrow] (and2) -- node[above, smalllabel] {\makecell{DERIV\_DOM\_\\[-.4cm]TOTALREL}} (deriv_dom);
			\draw[arrow] (deriv_dom) -- node[above, smalllabel] {\makecell{SIMP\_SUB-\\[-.4cm]SETEQ\_SING}} (simp_hyp);
			\draw[arrow] (simp_hyp) -- node[right, smalllabel] {HYP} (success2);
		\end{tikzpicture}
	}
	\caption{Application of Inference and Rewrite Rules to Prove a PO}
	\label{fig:ruleapplications}
\end{figure}
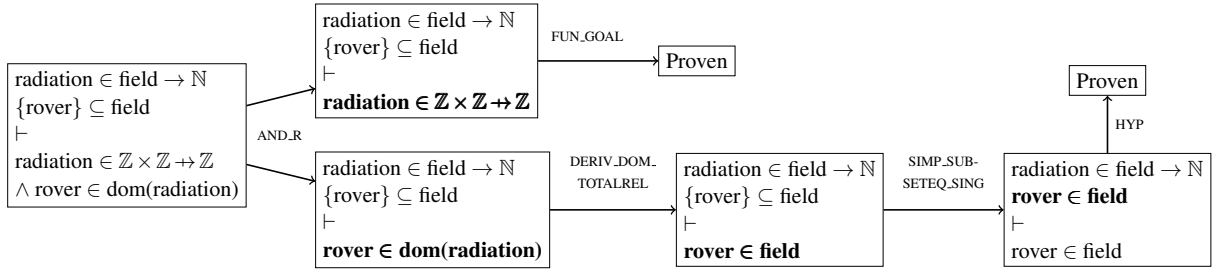

We use an example PO from an Event-B mars rover model%
\footnote{Available at \url{https://stups.hhu-hosting.de/models/sequent_prover}.} that will be used throughout the next chapters.
A safety invariant states that the rover must not be exposed to excessive radiation, expressed by
 \texttt{radiation(rover)} $\leq \alpha_{max}$, where 
  \texttt{radiation} assigns a radiation value to a position.
For this invariant, a well-definedness (WD) PO is created, as function application is only defined
  if \texttt{radiation} is actually a function
   and \texttt{rover} is in the domain of \texttt{radiation}.
A complete proof of the WD PO is depicted in \Cref{fig:ruleapplications}.
For brevity, 
we omit hypotheses that are not relevant for the proof.
Starting from the original goal on the left of \Cref{fig:ruleapplications}, the conjunction is split into two subgoals by applying the \texttt{AND\_R} rule.
To discharge the first resulting sequent at the top, we apply the \texttt{FUN\_GOAL} rule following from the first hypothesis. 
%
For the next sequent, 
we apply a rule for total relations and functions (\texttt{DERIV\_DOM\_TOTALREL}) to remove the domain operator $dom$ from the goal.
The second hypothesis contains a singleton set and can be simplified using the \texttt{SIMP\_SUBSETEQ\_SING} rule,
 after which we can discharge the goal by using the \texttt{HYP} rule.

\section{State-Based Representation of Proof Obligations}

\prob{}~\cite{leuschel2023prob} 
 is an animator and model checker for high-level formal models. 
It contains a constraint solver and interpreter for B and Event-B.
The XTL mode of \prob{}~\cite{xtl_iclp} allows one to circumvent the interpreter and directly
  load and then animate Prolog files that specify labelled transition systems.\footnote{Detailed documentation at: \url{https://prob.hhu.de/w/index.php?title=Other_languages}.}
The Prolog files define the ternary transition predicate \mintinline{prolog}{trans(Label,StateBefore,StateAfter)}, where the states of a transition system are represented as Prolog terms.
The initial states are provided using the \mintinline{prolog}{start(State)} predicate.
State properties can be specified with \mintinline{prolog}{prop(State,Property)}.
Here, we use them for better comprehensibility only. 
In the following, we use \prob{}'s XTL mode to turn the Event-B proof rules into an animatable transition system.

\begin{wrapfigure}[6]{r}{0.55\textwidth}
	\captionof{listing}{\centering Simple XTL Specification of a Button}
	\vspace{-.25cm}
	\hrule
	\vspace{-.1cm}
	\scriptsize
	\begin{minted}{prolog}
 start(button(on)).
 trans(toggle_button,button(X),button(Y)) :- toggle(X,Y).
 prop(button(X),'='(button,X)).
 toggle(on,off).
 toggle(off,on).
	\end{minted}
	\vspace{-.15cm}
	\hrule
	\label{lst:xtl_example}
\end{wrapfigure}

The code in \Cref{lst:xtl_example} demonstrates the XTL functionality for a simple button.
Initially, the button is \textit{on} and can be toggled to the other state via a transition.
The property shows the current state of the button.

For our sequent prover, the state representation is slightly more complex.
An example is shown in \Cref{lst:po_prolog_example}, illustrating the Prolog representation of the start state for the PO discussed in \Cref{sec:background}.
In general, a state has the form \mintinline{prolog}{state(sequent(SelHyps,Goal,Cont),Info)}.
The list of selected hypotheses is kept in the first entry \texttt{SelHyps}; the current \texttt{Goal} is the second entry.\footnote{Formulas are represented as a normalised abstract syntax tree. The term `$\$$' denotes an identifier.}
Only selected hypotheses are visible for proof rules; these can always be deselected and vice versa to hide those that are not required for a particular proof.
The \emph{continuation} \texttt{Cont} is either \texttt{success} or is the next still unproven sequent (which has another continuation itself).
In the initial state, there is exactly one open goal, so the continuation is \texttt{success}.
If the current state becomes \texttt{success}, then the proof has succeeded and there are no more open goals to show.
The \texttt{Info} field attached to the state is a list of additional information, including the deselected hypotheses and details about enumerated or deferred sets for typing.
%
\begin{mintlisting}[h]
    \vspace{-.2cm}
    \captionof{listing}{\centering PO proven in \Cref{fig:ruleapplications} represented as State Term}
    \vspace{-.25cm}
    \hrule
    \vspace{-.1cm}
    \footnotesize
    \begin{minted}{prolog}
 start(state(sequent(
     [member('$'(radiation), total_function('$'(field),'NATURAL')),
      subset(set_extension(['$'(rover)]),'$'(field))],
     conjunct(member('$'(radiation),
                     partial_function(cartesian_product('INTEGER','INTEGER'),'INTEGER')),
              member('$'(rover),domain('$'(radiation)))),
     success),Info),[description(Label)]).
    \end{minted}
    \vspace{-.15cm}
    \hrule
    \vspace{-.2cm}
    \label{lst:po_prolog_example}
\end{mintlisting}

Since \prob{} is not yet capable of \emph{generating} proof obligations (POs)\footnote{The WD prover can already compute WD conditions, which we utilise later in this article.}, we use the \rodin{} platform for generation of POs for Event-B specifications.
Using the \prob{} \rodin{}-Plugin, the generated POs can be exported as one file containing all obligations related to the selected machine (or context) as Prolog facts of the form \mintinline{prolog}{disprover_po(Label,Context,Goal,AllHyps,SelHyps,Status)}.
This format was originally designed for the \prob{} Disprover \cite{krings2015failure}.
Hence, we can easily reuse it to load and convert the POs for our sequent prover within \prob{}.
The first entry is the label of the PO, which distinguishes POs according to a systematic naming scheme.
For invariants, 
it consists of the event name, 
the label given to the invariant and 
the category of the PO 
(\emph{event/invariantlabel/INV}).
The \texttt{Goal} is represented as an \emph{untyped} abstract syntax tree (AST) as are all available hypotheses in \texttt{AllHyps} and pre-selected ones in \texttt{SelHyps}.
\texttt{Status} contains the \rodin{} proof status. 
The \texttt{Context} is used to extract user-defined types (Event-B sets), which are required for certain type rewriting rules. 

The Prolog representation of the goal and the hypotheses is normalised according to \prob{}'s WD prover~\cite{leuschel2020fast}.
Using the same format as the WD prover facilitates the integration of the proof rules into the \prob{} core.
Then, the normalised hypotheses are combined with the goal into a term that represents the initial \emph{state} of the PO as shown in \Cref{lst:po_prolog_example}.
Finally, all POs of a machine 
are made available as initial states together with their corresponding label \mintinline{prolog}{start(InitState,[description(Label)])},
where the optional second argument is for the transition property explained in the following section.




\section{Proof Rules as a Transition System}\label{sec:prolog_proof_rules}


To define how proof rules transform a sequent into another or more successor sequents, the transition predicate that was introduced in the previous section can be used.
Take for example the 
  rule \texttt{SIMP\_SUBSETEQ\_SING} 
  we applied in \Cref{sec:background}. 
It encodes $\{ E\}  \subseteq  S  \equiv  E \in  S$, which we can implement in Prolog and XTL as shown in \Cref{lst:simp_rule_example}.\footnote{We make minor modifications of the source code here to ease understanding.}
\begin{mintlisting}[h]
    \captionof{listing}{\centering Implementation of Simplification and Rewrite Rules}
    \vspace{-.25cm}
    \hrule
    \vspace{.1cm}
    \footnotesize
    \begin{minted}{prolog}
   trans(simplify_goal(RuleName),state(sequent(Hyps,OldGoal,Cont),Info),
                                 state(sequent(Hyps,NewGoal,Cont),Info)) :-
         simp_rule(OldGoal,NewGoal,RuleName,Hyps).
   trans(simplify_hyp(RuleName),state(sequent(OldHyps,Goal,Cont),Info),
                                state(sequent(NewHyps,Goal,Cont),Info)) :-
         select(Hyp,OldHyps,NewHyp,NewHyps), simp_rule(Hyp,NewHyp,RuleName,OldHyps).
   ...
   simp_rule(subset(SetA,S),member(A,S),'SIMP_SUBSETEQ_SING',_) :- singleton_set(SetA,A).
   simp_rule(not_equal(L,L),falsity,'SIMP_MULTI_NOTEQUAL',_).
    \end{minted}
    \vspace{-.15cm}
    \hrule
    \vspace{-.25cm}
    \label{lst:simp_rule_example}
\end{mintlisting}
The first two clauses group together several simplification rules, which we apply either to the goal or to a specific hypothesis.
That is, there are many other clauses for {\tt simp\_rule}, two of which we show in \Cref{lst:simp_rule_example}.
This separation allows us to express a formal rule intuitively as one Prolog clause, with the original formula as the first argument, the rewritten formula as the second argument, and additional side conditions forming the body.

In \Cref{lst:example_proof_rules_prolog}, we show a selection of other rules.
%
%
\begin{figure}[h]
    \vspace{-.15cm}
    \captionof{listing}{\centering Prolog Rules Used for the Proof in Figure \ref{fig:ruleapplications}}
    \vspace{-.25cm}
    \hrule
    \vspace{-.1cm}
    \small
    \begin{minted}[escapeinside=@@]{prolog}
    simp_rule(domain(R),Dom,'DERIV_DOM_TOTALREL',Hyps) :-
        member(member(R,RType),Hyps), is_rel(RType,total,Dom,_).
    is_rel(total_relation(Dom,Ran),total,Dom,Ran).
    ...
    trans(and_r,state(sequent(Hyps,conjunct(G1,G2),Cont),I),
                state(NewSequent,I)) :-
         op_to_list(conjunct(G1,G2),ListOfConjuncts,conjunct),
         add_goals(Hyps,ListOfConjuncts,Cont,NewSequent).
    trans(Rule,state(sequent(Hyps,Goal,Cont),I),state(Cont,I)) :-
         axiom(Rule,Hyps,Goal).
    trans(Rule,state(sequent(Hyps,Goal,Cont),I),state(Cont,I)) :-
         axiom_with_info(Rule,Hyps,Goal,I).
    ...
    axiom(hyp,Hyps,Goal) :- member_hyps(Goal,Hyps).
    ...
    axiom_with_info(fun_goal,Hyps,member(F,partial_function(Ty1,Ty2)),Info) :-
        type_expression(Ty1,Info), type_expression(Ty2,Info),
        member(member(F,FType),Hyps), is_fun(FType,_,_,_).
    \end{minted}
    \hrule
     \vspace{-.3cm}
    \label{lst:example_proof_rules_prolog}
\end{figure}

\begin{itemize}
    \item For example, there are more complicated simplification rules like \texttt{DERIV\_\-DOM\_TOTALREL}, where the rule needs access to all hypotheses.
    Note that in \mintinline{prolog}{member(member(R,RType),Hyps)}, the inner \emph{member} represents the AST node of the membership operator and the outer one checks for Prolog list membership.
    \item We show the encoding of the \texttt{AND\_R} rule, which decomposes the goal \mintinline{prolog}{conjunct(G1,G2)} into its conjuncts.
    The current goal is replaced with the leftmost conjunct and the other conjuncts are added to the continuation (this is done by \texttt{add\_goals} in line 8 of \Cref{lst:example_proof_rules_prolog}).
    \item We have \emph{axioms} as terminal proof steps,
    which close a proof branch without generating successor sequents, e.g. \texttt{HYP} and \texttt{FUN\_GOAL}.
\end{itemize}

\texttt{HYP} is applied whenever the current goal is already contained in the hypotheses, i.e., when unification succeeds.
One must consider that Prolog unification is insufficient for associative-commutative operations to determine equality \cite{lincoln1989adventures}.
Expressions such as \mintinline{prolog}{add(add(x,y),z)} and \mintinline{prolog}{add(x,add(y,z))} or \mintinline{prolog}{add(add(z,y),x)} cannot be unified directly.
Instead, in \texttt{member\_hyps} ground terms are then normalised into special lists to remove nesting, and compared by sorting.
Moreover, it is taken into account that hypotheses may be stronger than the goal.
For example, \texttt{HYP} can be applied for $y < x \vdash x \geq y$.

The applicability of \texttt{fun\_goal} depends on the goal having a particular form.
\texttt{type\_expression} facts accept base types in the first argument directly, whereas composite type expressions are validated recursively.
The second argument (\texttt{Info}) is used to look up 
custom data types 
in the list of meta-information.
To make type information available, including user-introduced types, the hypotheses are type-checked at the beginning of the proof and the resulting information is stored in \texttt{Info}.

Rewrite rules can be applied to both top-level expressions and subterms by recursively descending into the structure of a term and attempting to apply a simplification rule to its arguments.
That is why \mintinline{prolog}{domain('$'(radiation))} in \mintinline{prolog}{member('$'(rover),domain('$'(radiation)))} can be simplified to \mintinline{prolog}{'$'(field)} (recall \Cref{fig:ruleapplications}).

%

When applying particular rules, new identifiers need to be introduced, for instance, when rewriting $S \neq \emptyset$ to $\exists x \cdot x \in S$ with a fresh identifier $x$.
To achieve this, a list of existing identifiers in an expression is computed, and a new identifier is generated that does not appear in that list.


\paragraph{Transition Descriptions.}

\prob{} allows to add \emph{dynamic} transition properties 
depending on the parameter values and the state 
in which the transition is available. 
In our case, we use this feature to add user-friendly transition descriptions 
by adding a property \texttt{description/1}.
These are used 
to display the transition descriptions in \probtwoui{} or in the HTML exports (see, e.g., \Cref{fig:state_vis_curr,fig:proof_html_rodin}).



\paragraph{Rules with User Input.}


Certain inference rules require user input, such as when adding new hypotheses or instantiating free identifiers of a universally quantified variable:
\[\texttt{FORALL\_INST}\qquad \frac{\textbf{H} \;\;\vdash \; \textit{WD}(E) \qquad \textbf{H} , [x \bcmeq E]\textbf{P} \;\;\vdash \;\; \textbf{G}}{\textbf{H}, \forall x \qdot \textbf{P} \;\;\vdash\;\; \textbf{G}}\]

For this, we use \emph{symbolic} transitions specified by the predicate \texttt{symb\_trans/3} in a similar way as for \texttt{trans/3}.
However, symbolic transitions are not automatically evaluated during animation and can only be invoked manually. 
In the case of a universal quantifier, the user must provide the index of the universally quantified hypothesis 
to be instantiated, as well as a B expression \texttt{E} 
for the instantiation.
In addition, \texttt{E} must be proven to be well-defined.
We use \prob{}'s WD analyser to compute 
\texttt{WD(E)}.

\paragraph{External Provers.}

As in \rodin{}, it is possible to interface to external provers, such as SMT solvers, namely Z3, the Atelier-B ML/PP provers, the \prob{} 
 constraint-based disprover, and the specialised \prob{} prover for WD conditions.
The interfaces to the SMT solvers and the Atelier-B provers are already part of \prob{} and are reused.
Calls to external provers can be time-consuming, so they are implemented as symbolic transitions to prevent all provers from being called each time a state is explored.

\section{Proving with \prob{}} 

\begin{wrapfigure}[12]{r}{0.35\textwidth}
	\centering
	\vspace{-.4cm}
	\includegraphics[width=\linewidth]{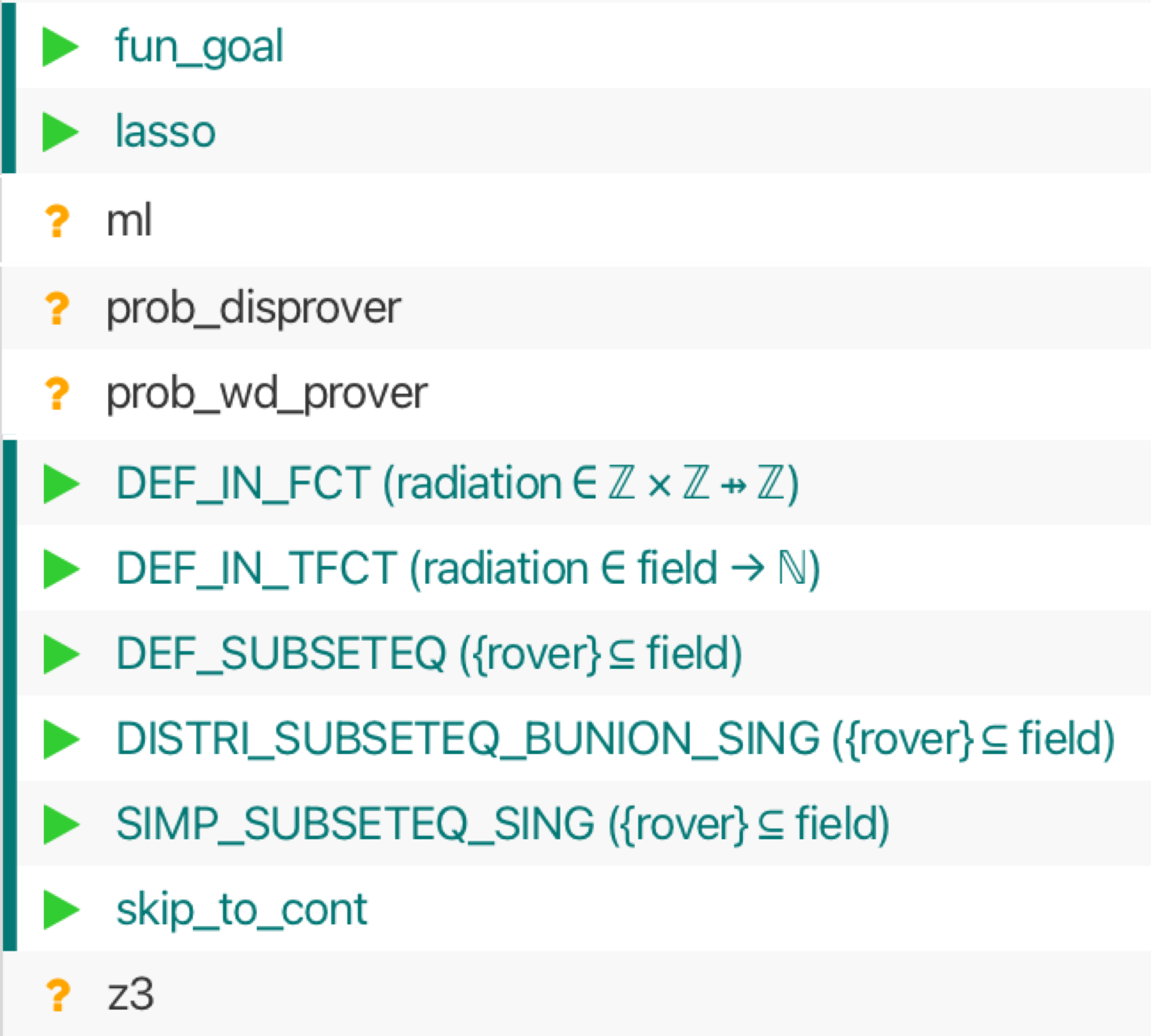}
	\caption{\centering Rules as Transitions in the \prob{} Animator} 
	\label{fig:animator}
\end{wrapfigure}
We integrated the proof rules into the \prob{} core, enabling direct loading of PO files exported with the \prob{} Disprover plugin from \rodin{}.
The animator automatically enters the new \textit{sequent prover} animation mode, providing access to the proof transitions (\Cref{fig:animator}).
With the animator, we obtain a user interface to the prover and full access to its debugging capabilities within the \prob{} tooling, e.g. via \probtwoui{}~\cite{prob2_ui}.
We present new export options for proofs created with \prob{}.\footnote{Examples can be explored online at \url{https://stups.hhu-hosting.de/models/sequent_prover}.}
The aim is twofold: to make the proof more comprehensible for the user and to enable it to be exported to other tools.


\subsection{Save and Replay Proof Trace}\label{sec:probtrace}

A proof created by \prob{} corresponds to a sequence of applied proof rules, i.e. a \emph{linear} trace of transitions.
Such a trace can be saved as a JSON file that contains all relevant steps, parameter values, and state terms for later replay with \prob{}.
This also enables interactive trace replay \cite{rodin_2025_interactive_trace_replay}, allowing a gradual and controlled replay of a proof.
This can be advantageous in several scenarios, for instance, when proofs have to be refactored after small changes to the Event-B model, where only a few rules within a proof trace need to be replaced.
It can be employed to replay a trace created for an abstract Event-B model on the refined model (or vice versa), thereby \emph{refining the proof} without having to perform it from scratch.
Moreover, it is possible to save a trace of a partial proof and continue it later.

\subsection{Visualisation}\label{sec:visualisation}

\prob{} integrates various visualisation mechanisms.
These include graph visualisations created with Graphviz~\cite{gansner2009drawing} and state visualisations controlled either by an animation function or by \visb{}~\cite{visb}, which enables domain-specific visualisations based on SVG graphics.
We use these interfaces to generate comprehensible proof representations and allow users to control the proof by clicking in the visualisation.

\paragraph{Interactive State Visualisation.}

To represent the current proof state during animation, we use \prob{}'s state visualisation feature.
The visualisation is grid-based, where each cell can contain text or an image.
The cell contents are determined by a special Prolog predicate representing the animation function, with the current state as input. 

\begin{wrapfigure}[9]{r}{0.62\textwidth}
	\centering
	\includegraphics[width=\linewidth]{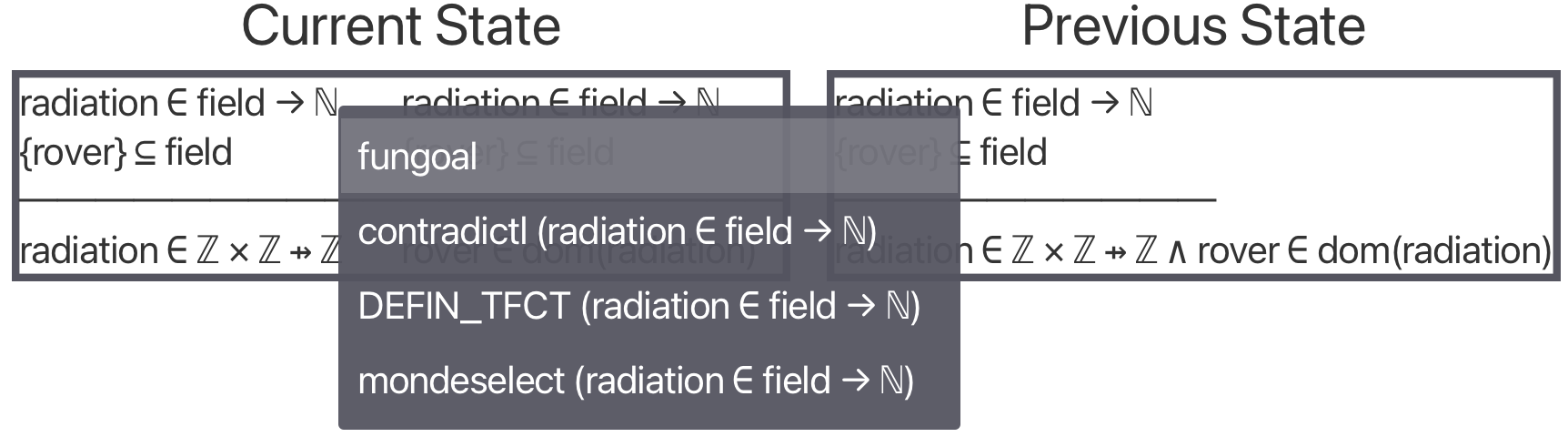}
	\caption{\centering Visualisation of the Current and Previous Proof Sequent with Applicable Rules for One Row}
	\label{fig:state_vis_curr}
\end{wrapfigure}

For our sequent prover, the animation function deconstructs the state into the current state and its continuations, 
displaying the current sequent in the leftmost column, with its continuations in the subsequent columns. 
In addition, the previous state is displayed next to the current state. 
See \Cref{fig:state_vis_curr} for a visualisation of our example introduced in \Cref{sec:background} in \probtwoui{}.
Selected hypotheses appear 
above the line, while the current goal is shown below.
The predicates are rendered in a readable manner using the pretty-printer for B expressions of \prob{}.
Using its Unicode mode produces a representation that closely resembles mathematical notation.
Additionally, a visualisation of the previous sequent is provided for comparison of the changes made by the recent transition.
The visualisation is interactive, allowing the user to select a row in the current sequent by right-clicking on it.
%
%
The applicable rules for the selected row can be accessed via a pop-up menu, using the associated transition descriptions (cf. \Cref{sec:prolog_proof_rules}).
In comparison, \rodin{} allows direct application of manual rewrite rules by clicking on symbols highlighted in red within a formula.
%
%
Here, the currently applicable rules can be selected via the operations view, and the already performed steps can be inspected in the history.
The state visualisation is accompanied by the state view, which lists the current hypotheses and the goal.

\vspace{-.1cm}
\paragraph{Proof Tree.}

With the \prob{} animator, proofs (or proof attempts) are represented as traces leading to the current state (cf. \Cref{sec:probtrace}), with open proof nodes represented by continuations in that state.
However, the user can also skip to a continuation during the animation, so that a trace does not itself provide a proof tree.
Hence, we implemented a conversion from a proof \textit{trace} to a \textit{tree} representation.
This can then be visualised using the interface from \prob{} to Graphviz, and is used for exporting proofs 
(see \Cref{sec:bpr_rodin}).

Converting a linear proof trace into a proof tree involves creating nodes for each continuation, with edges representing the applied proof rules.
Each edge links a sequent to its new continuations 
resulting from the application of the rule.
When a rule discharges the goal, the corresponding edge connects to a node representing the successful proof for that branch.

\begin{figure}[t]
	\centering
	\begin{minipage}{0.62\textwidth}
		\begin{center}
			\includegraphics[width=\textwidth]{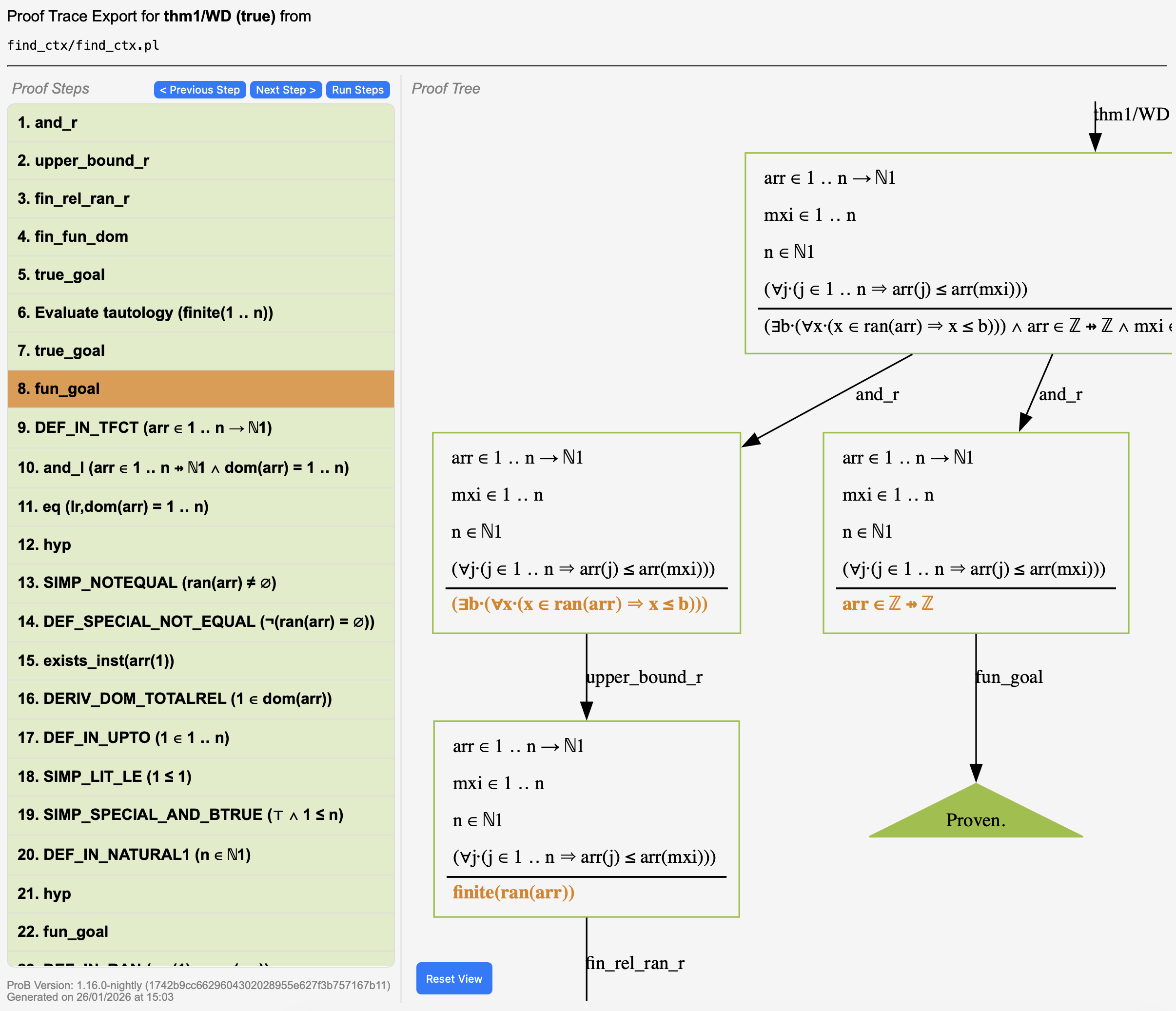}
			\caption{\centering Interactive HTML Proof Tree}
			\label{fig:proof_html_rodin}
		\end{center}
	\end{minipage}
	\hfill
	\begin{minipage}{0.37\textwidth}
		\begin{center}
			\includegraphics[width=\textwidth]{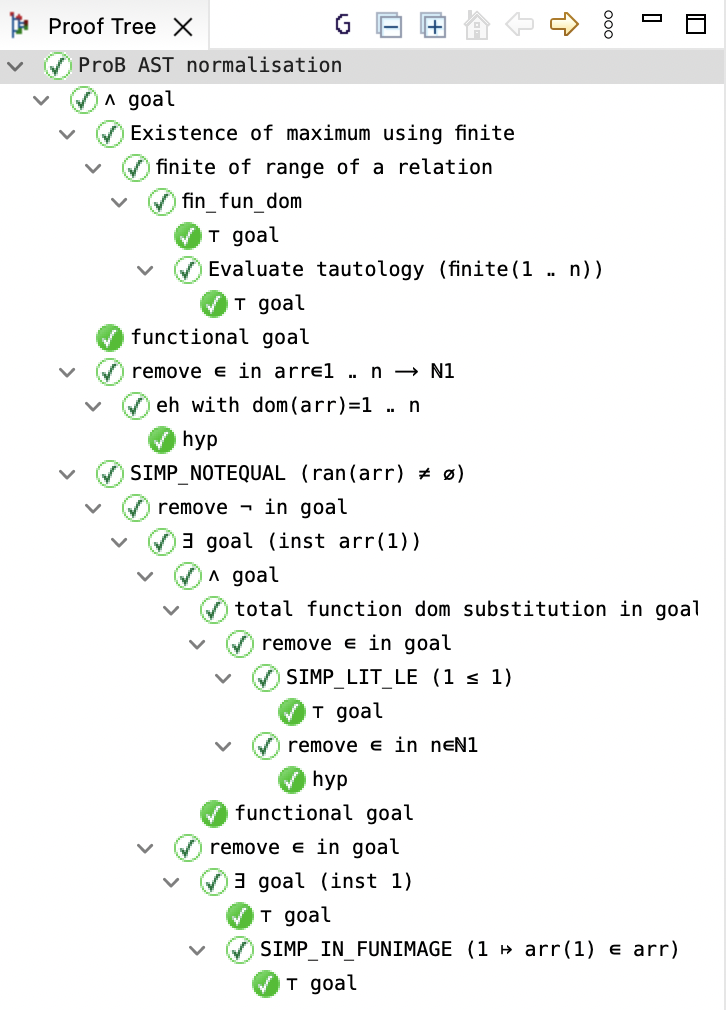}
			\caption{\centering Proof replayed by the Sequent Prover in \rodin{}}
			\label{fig:proof_bpr_rodin}
		\end{center}
	\end{minipage}
\end{figure}

We integrated a visualisation of the proof tree in 
\prob{}'s graph visualisations, easily accessible via the user interfaces (e.g. CLI, \probtwoui{}).
It allows users to inspect the current state in the context of the proof tree.
An example is shown in \Cref{fig:proof_html_rodin}. 
By interfacing with Graphviz, the visualisation can be exported in multiple other file formats, including PDF, SVG, PNG, or as raw DOT file.

%

\vspace{-.1cm}
\paragraph{Interactive HTML Documents.}

Often, proofs are hard-wired to a specific tool and cannot be exported easily.
For instance, \rodin{} stores proofs as raw XML data in .bpr files, limiting the ability to share the proof outside of the tool.
In contrast, \prob{} allows to export traces of formal models with state visualisations as standalone HTML documents.
However, the existing export options are not ideally suited for proofs.
To address this, we developed a new HTML export based on the proof tree. 
Note that the HTML file generation is also implemented in Prolog.
Hereby, we generate an SVG visualisation of the proof tree using Graphviz.
Our visualisation highlights which parts of the sequent have changed after a proof step.
Users can manually adjust their view by zooming and panning.
In addition, we list the applied proof rules, i.e. the edges of the proof tree, next to the visualisation. 
Users can interactively click on proof steps to enlarge the relevant parts of the tree and navigate through it step by step.

Figure \ref{fig:proof_html_rodin} shows a screenshot of an example.
Rule 8 has been selected, corresponding to the right branch in the enlarged section of the visualisation.
The green triangle indicates that the applied proof rule (\texttt{fun\_goal}) has discharged the goal, and this branch of the proof tree is complete. 
The left branch has not yet been proven, and can be explored either by panning in the visualisation or by navigating to the next step in the step view.

In general, the export allows a detailed and interactive inspection of the applied proof steps and the proof tree itself.
It is particularly useful for teaching because adjacent proof nodes are displayed clearly in the proof tree, unlike in the visualisation in \rodin{}.
Also, only a web browser is required to inspect a proof. 

\subsection{Export for Replay with Rodin}\label{sec:bpr_rodin}

In addition to the human-readable HTML outputs, 
we developed an XML export for proof replay within \rodin{}.
\rodin{} stores proof trees, along with the applied steps and the associated Java reasoners, in BPR files, which are based on XML. 
For our examples, complete BPR files can be found on the web server.

To export a proof, we generate a BPR file in Prolog based on our proof tree (cf. \Cref{sec:visualisation}), transforming proof nodes into continuations
and edges into proof rule applications.
In addition, the required Java reasoners 
are specified, which are called by \rodin{} during the replay of the proof.

If the \rodin{} reasoners can replay a proof created with \prob{}, we obtain additional \emph{validation of the proof}, enabling \prob{} to be used as a \emph{second chain}. 
Users can also export a partial proof from \prob{} to \rodin{}, yielding a more detailed proof tree in \rodin{} (see \Cref{fig:proof_bpr_rodin}), instead of nodes that apply several steps at once.
For example, when the ML (Mono Lemma) prover from \atelierb{} discharges a goal in \rodin{}, the applied tactics remain completely hidden from the user (compared to \Cref{fig:proof_bpr_rodin}, there would be just a single success node with no further details).
Furthermore, unlike a pure \rodin{} proof, our export includes the precise rule names as comments. 

However, there are still a few limitations in the mapping of our proof rules to the \rodin{} reasoners. 
Specifically, many of the automatic rewrite rules cannot be translated from our detailed proof steps, since \rodin{} applies most of the simplification rules recursively in one step to all hypotheses and conceals them behind the message ``simplification rewrites''. 
For these rules, we generate individual proof steps that are marked as ``reviewed'' (the step is actually reviewed by \prob{}), enabling unverified replay in \rodin{}.
The same approach is used 
for our custom rules without an equivalent in \rodin{}.
This can be observed in~\Cref{fig:proof_bpr_rodin}, where we used the \prob{} sequent prover to apply the rule \texttt{FIN\_FUN\_DOM}, which is not yet implemented in \rodin{}.
While the replay succeeds using the aforementioned review, it is not possible to perform the same proof with \rodin{} standalone. 

%




\subsection{Automated Proving}

\paragraph{Model Checking}
One can use \prob{}'s model checker as an automatic prover.
This can be useful to find proofs, but performance-wise this is not ideal:
  the model checker will store every reached proof state and will always compute all possible
  applications of proof rules for an explored proof state.
  

\paragraph{Auto Prover} \label{auto-prover}
Luckily, search is one of the strengths of Prolog.
As such, it is easy to encode a custom proof search on top of our Prolog encoding of the Event-B proof rules.
We have thus implemented a first automatic prover, using iterative deepening search with a few simple heuristics.
The heuristics describe ``eager'' proof rules that should always be applied (deterministically first) like {\tt AND\_R} and {\tt AND\_L}
 to decompose conjunctions in the proof goal and in the hypotheses.
This prover is already useful to automatically find short proofs (it can be run automatically in our animator interface).

In the future, we wish to encode more clever heuristics and enforce ordering of the proof rules.
By using partial evaluation, we expect to obtain a very fast, rule-based prover for Event-B which ideally will
surpass the performance of current provers like the ML prover from \atelierb{}.

Indeed, \atelierb{} uses the {\em theory language} to express proof rules, which can be viewed as domain specific
logic ``programming language'' tailored to B and proof.
While \atelierb{} comes with a custom developed compiler -- the {\em Logic Solver} -- it seems like it cannot compete with state-of-the-art Prolog compilers
 (see \cite{leuschel2020fast}, in particular Section 6).
This suggests that, at least in principle, our Prolog encoding could lead to a faster automated prover.

\section{Comparison with Java Implementation}

Using Prolog for verification tools has been advocated by Leuschel~\cite{Le08_233,DBLP:journals/corr/abs-2008-02933}.
In this section we compare our Prolog implementation with the existing Java implementation of the Event-B proof rules.

The \rodin{} platform has now been developed for well over 20 years.
Our sequent prover was developed since summer 2025 during a 20 ETCS credit points project by the first author (i.e., about 600 hours of work),
with additional developments by the other authors (notably on proof visualisation, export and auto-proving). 
 
Table~\ref{tab:comparison}  provides a brief overview of the Java source code of the sequent prover
 compared to our Prolog implementation.%
\footnote{
The columns SLOC (Source Lines of Code) was generated using David A. Wheeler's `SLOCCount' tool. The tool also computes an estimated number of person years: 6.51 for the Rodin sequent prover and 0.79 for our prover. At least the latter number is quite close to the real development effort.}
Note that the intention of this table is to give a rough estimate
 of the size of the implementation.


\begin{table}[ht]
	\centering
	\caption{Statistics of Java and Prolog Implementation}
	\vspace{-.2cm}
    \begin{tabular}{ll|rrr|rr}
        \hline
        Tool & Language & \multicolumn{3}{c|}{Code for Sequent Prover} & \multicolumn{2}{c}{Missing Rules} \\
         &   &  Files & LOC & SLOC  & Rewrite & Inference \\
        \hline
        \rodin{} & Java & 320 & 50182 & 27580  & 55/536 & 16/123 \\
        \prob{} & Prolog & 4 & 4208 & 3705  & 4/536 & 8/123 \\
        \hline
    \end{tabular}
	\label{tab:comparison}
	\vspace{-.2cm}
\end{table}


 %

Our implementation covers more proof rules (not counting the additional proof rules taken from~\cite{Abrial10} as well as few additional ones for Peano arithmetic).
In terms of code size there is an order of magnitude difference when counting lines of code (LOC) and a bit less when counting source lines of code (SLOC, i.e., disregarding comments and whitespace).

In general, new proof rules can be added easily to our Prolog implementation, sometimes requiring a single clause
as opposed to a new class in Java. 
Consider for example the \texttt{DERIV\_DOM\_TOTALREL} rule we have used in \Cref{sec:background}.
The Prolog clause just expresses the logical condition when the domain of a total relation can be rewritten (cf. top of \Cref{lst:example_proof_rules_prolog}).
Beyond this logical core, the Java implementation involves the construction of many objects and various additional checks.
It also distinguishes explicitly between rewriting in goals and hypotheses, which we do not have to do in Prolog.
This leads to a class of 257 LOC 
 compared to one clause in Prolog (which facilitates maintenance). 
%
%
It is also indicative (and maybe not surprising) that the Java implementation is still missing 71 proof rules,
 20 years on. 
While the exact reason for this is unclear to us,
 it at least suggests that non-trivial development effort is required to add new proof rules in Java.

Earlier work by Leuschel~\cite{leuschel2020fast} has shown several orders of magnitude difference between another Prolog prover and running the \rodin{} provers.
The prover presented there, however, is a dedicated prover for well-definedness
 and is not complete (it typically does not do case distinctions to avoid exponential blow-up).
Our prover presented in this paper is not yet tuned for performance, and the tactics of the auto-prover need to be fully implemented.
Therefore, we have not yet tested whether the Prolog implementation of the proof rules or the Java counterpart is more efficient.
In the future, we intend to re-use the hypothesis management from \cite{leuschel2020fast},
 along with partial evaluation to obtain an efficient, yet extensible and maintainable prover.
We also plan to incorporate custom proof rules written by users using \rodin's theory plugin~\cite{DBLP:conf/birthday/ButlerM13}.


\section{Teaching Aspects}


Modelling the mars rover was a project in our university course on safety-critical systems, where students learn the development and verification of reliable software using the formal B method.
The system is considered safety-critical due to financial concerns, as damage to or loss of the rover would result in significant costs.
Using the interactive sequent prover can provide insight into the reasoning behind proofs that are otherwise discharged automatically in \rodin.
The existence of different ways to provide a successful proof 
is another reason why interactive exploration is useful for teaching.

With the \prob{} sequent prover, we could prove all proof obligations of a machine 
from the self-developed model.
On the contrary, proving preservation of the same invariants was rather challenging in \rodin{}, although we knew why the invariants must hold after execution of the events.
Some proofs went through after clicks through symbols and the repeated pruning of proof branches, which did not help to gain a clear understanding of why the proof succeeded.
This experience allows us to assume, that using our prover improves the user experience from an educational point of view, as the effect of each applied rule is visible.
Together with our HTML proof export, students can focus on proof rules instead of specific tool buttons and can examine the interaction of the rules more systematically. 


For future work, we are considering ``gamification'' of the animation to give more insights, for example by using heuristics to provide additional guidance on promising rules in the current state, and restricting the set of available rules. 
There are similar approaches, for instance, Hendriks et al.~\cite{hendriks2010teaching} developed an interactive web interface for teaching logic 
based on the proof assistant Coq.
Cumbor et al.~\cite{CumborStoddartDunne2010} conducted a ``proof game'' approach for inference rules, 
which they report as being popular among students.
Another example of proof gamification is Kevin Buzzard's Natural Number Game for Lean\footnote{\url{https://adam.math.hhu.de/\#/g/leanprover-community/nng4}}, which introduces the writing of formal proofs in Lean by interactive exercises.
A recent study analysed the interaction of undergraduate students with the tool \cite{iannone2025feels}.


\section{Related Work}

The work by Grieu et al.~\cite{grieu2025translating} translates \rodin{} proofs to the TLA$^+$ Proof System (TLAPS). 
They encountered the issue that \rodin{} applies many simplification rules together in a single step,
 making a translation challenging (as described in \Cref{sec:bpr_rodin}).
Our more fine-grained proof trees would solve this problem (and Prolog would be a good language
 to encode such translation rules).

The Why3 Plug-In for \rodin{} \cite{DBLP:conf/asm/IliasovSAR16} translates proof obligations into Why3,
but proofs are not translated back to Event-B proofs.
Here the Dedukti research project~\cite{DBLP:journals/corr/abs-2311-07185} is more relevant: it has the goal to share proofs across systems.
Some first works try to translate B/Event-B proofs to the {\(\lambda\)}{\(\Pi\)}-Calculus~\cite{stolze:hal-04398119,DBLP:conf/zum/Grieu24}.
Our work helps understand, visualise and check proofs, but is based on a translation to a programming language (Prolog) rather
 than a proof calculus.


As already discussed in Section~\ref{auto-prover},
 Atelier-B uses a domain-specific ``theory language'' to express the proof rules.
The execution engine (krt) written in C seems to have a performance far below
that of mature Prolog engines (see \cite{leuschel2020fast}). 

There are many successful interactive provers, a.k.a. proof assistants
 such as Coq, PVS, Lean and Isabelle/HOL.
A translation of Event-B to Isabelle was studied by Schmalz~\cite{DBLP:phd/basesearch/Schmalz12}.
The work was never completed to a point where one could use Isabelle for arbitrary proof obligations.

LeanTAP~\cite{DBLP:journals/jar/BeckertP95,DBLP:journals/logcom/Fitting98}
is a theorem prover for first-order logic written in a few lines of Prolog.
Its performance is one inspiration for our work.
Another successful 
prover is Vampire~\cite{DBLP:journals/aicom/RiazanovV02}.
LeanTAP and Vampire are based mainly on resolution, while in this work we have encoded many different B-specific proof rules.


One interesting question is whether Prolog or a term rewriting system such as Maude
 is more appropriate to encode our proof rules.
Prolog supports full unification, but only at the top-level of a term,
 while term rewriting systems support matching and rewriting of subterms.
For most of our inference rules, unification at the top-level was sufficient, e.g., for {\tt AND\_R} in Section~\ref{sect-with-andr}
 which looks for an outer conjunction in a proof goal.
The simplification rules, however, require traversing the proof goal or the hypotheses for matching subterms.
This traversal is encoded in one separate predicate; the simplification rule encodings again match at the top-level.
We are not aware of any efforts to encode B proof rules in term-rewriting systems.
The BMaude tool~\cite{DBLP:journals/corr/abs-2108-07878} 
provided support for executing a subset of classical B in Maude, but according to past experiments it was considerably slower than \prob{}.%
\footnote{E.g., 8.6 seconds vs 0.9 seconds to model check a mutual exclusion B model with 8008 states. Experiment conducted by the last author in August 2018.}
The newer EventB2Maude tool~\cite{DBLP:conf/nfm/OlarteORR25} supports Event-B and provides some inference rules for probabilistic reasoning.



\section{Conclusion} 

In summary, we have implemented a new sequent prover for the Event-B formal method.
The implementation covers more than 600 proof rules, and after 9 months of development it covers more
 of the standard proof rules than the previous Java implementation after 20 years.
The code is an order of magnitude more compact and should be easier to maintain and extend in the future.
By integrating the proof rules into \prob{}, we have managed to use
 the animation features of \prob{} for interactive proving, proof replay and proof repair.
New visualisation and HTML export features have been added to \prob{},
 enabling better human understanding of proofs.
A preliminary automated prover is implemented and proofs can be re-exported
 and double-checked in the \rodin{} platform. 




\condtext{}{
In the future, we plan to integrate proving features more deeply into the \probtwoui{}, providing access to the auto prover and immediate start of a proof without having to export the POs from \rodin{} first.
By using the AI interface of \simb{} (\cite{validation-rl-safety-shields}), we could already perform AI-controlled proofs, given a suitably trained AI. 
We aim to extend and adapt the rules for the classical B method and also generate POs in \prob{}.
For \prob{} in general, the rewrite rules could also be used for AST simplifications.


linking more external provers to ProB (Vampire/PVS/...)

 \simb{}~\cite{simb} (to encode tactics?), 
 interactive~\cite{simb_interactive} 
partial evaluation, obtain \prob{} WD prover?
}


\bibliographystyle{eptcs}
\bibliography{references}

\begin{thebibliography}{10}
\providecommand{\bibitemdeclare}[2]{}
\providecommand{\surnamestart}{}
\providecommand{\surnameend}{}
\providecommand{\urlprefix}{Available at }
\providecommand{\url}[1]{\texttt{#1}}
\providecommand{\href}[2]{\texttt{#2}}
\providecommand{\urlalt}[2]{\href{#1}{#2}}
\providecommand{\doi}[1]{doi:\urlalt{https://doi.org/#1}{#1}}
\providecommand{\eprint}[1]{arXiv:\urlalt{https://arxiv.org/abs/#1}{#1}}
\providecommand{\bibinfo}[2]{#2}

\bibitemdeclare{book}{bmethod}
\bibitem{bmethod}
\bibinfo{author}{Jean-Raymond \surnamestart Abrial\surnameend}
  (\bibinfo{year}{2005}): \emph{\bibinfo{title}{{The B-Book: Assigning Programs
  to Meanings}}}.
\newblock \bibinfo{publisher}{Cambridge University Press}.

\bibitemdeclare{book}{Abrial10}
\bibitem{Abrial10}
\bibinfo{author}{Jean-Raymond \surnamestart Abrial\surnameend}
  (\bibinfo{year}{2010}): \emph{\bibinfo{title}{Modeling in {Event-B}: System
  and Software Engineering}}.
\newblock \bibinfo{publisher}{Cambridge University Press}.

\bibitemdeclare{article}{rodin}
\bibitem{rodin}
\bibinfo{author}{Jean-Raymond \surnamestart Abrial\surnameend},
  \bibinfo{author}{Michael \surnamestart Butler\surnameend},
  \bibinfo{author}{Stefan \surnamestart Hallerstede\surnameend},
  \bibinfo{author}{Thai~Son \surnamestart Hoang\surnameend},
  \bibinfo{author}{Farhad \surnamestart Mehta\surnameend} \&
  \bibinfo{author}{Laurent \surnamestart Voisin\surnameend}
  (\bibinfo{year}{2010}): \emph{\bibinfo{title}{{Rodin: An Open Toolset for
  Modelling and Reasoning in Event-B}}}.
\newblock {\slshape \bibinfo{journal}{Int. J. Softw. Tools Technol. Transf.}}
  \bibinfo{volume}{12}(\bibinfo{number}{6}), p. \bibinfo{pages}{447–466},
  \doi{10.1007/s10009-010-0145-y}.

\bibitemdeclare{article}{DBLP:journals/corr/abs-2311-07185}
\bibitem{DBLP:journals/corr/abs-2311-07185}
\bibinfo{author}{Ali \surnamestart Assaf\surnameend},
  \bibinfo{author}{Guillaume \surnamestart Burel\surnameend},
  \bibinfo{author}{Rapha{\"{e}}l \surnamestart Cauderlier\surnameend},
  \bibinfo{author}{David \surnamestart Delahaye\surnameend},
  \bibinfo{author}{Gilles \surnamestart Dowek\surnameend},
  \bibinfo{author}{Catherine \surnamestart Dubois\surnameend},
  \bibinfo{author}{Fr{\'{e}}d{\'{e}}ric \surnamestart Gilbert\surnameend},
  \bibinfo{author}{Pierre \surnamestart Halmagrand\surnameend},
  \bibinfo{author}{Olivier \surnamestart Hermant\surnameend} \&
  \bibinfo{author}{Ronan \surnamestart Saillard\surnameend}
  (\bibinfo{year}{2023}): \emph{\bibinfo{title}{Dedukti: a Logical Framework
  based on the {\(\lambda\)}{\(\Pi\)}-Calculus Modulo Theory}}.
\newblock {\slshape \bibinfo{journal}{CoRR}} \bibinfo{volume}{abs/2311.07185},
  \doi{10.48550/ARXIV.2311.07185}.
\newblock \eprint{2311.07185}.

\bibitemdeclare{article}{DBLP:journals/jar/BeckertP95}
\bibitem{DBLP:journals/jar/BeckertP95}
\bibinfo{author}{Bernhard \surnamestart Beckert\surnameend} \&
  \bibinfo{author}{Joachim \surnamestart Posegga\surnameend}
  (\bibinfo{year}{1995}): \emph{\bibinfo{title}{{leanTAP}: Lean Tableau-based
  Deduction}}.
\newblock {\slshape \bibinfo{journal}{J. Autom. Reasoning}}
  \bibinfo{volume}{15}(\bibinfo{number}{3}), pp. \bibinfo{pages}{339--358},
  \doi{10.1007/BF00881804}.

\bibitemdeclare{inproceedings}{prob2_ui}
\bibitem{prob2_ui}
\bibinfo{author}{Jens \surnamestart Bendisposto\surnameend},
  \bibinfo{author}{David \surnamestart Gele\ss{}us\surnameend},
  \bibinfo{author}{Yumiko \surnamestart Jansing\surnameend},
  \bibinfo{author}{Michael \surnamestart Leuschel\surnameend},
  \bibinfo{author}{Antonia \surnamestart P\"utz\surnameend},
  \bibinfo{author}{Fabian \surnamestart Vu\surnameend} \&
  \bibinfo{author}{Michelle \surnamestart Werth\surnameend}
  (\bibinfo{year}{2021}): \emph{\bibinfo{title}{{ProB2-UI: A Java-based User
  Interface for ProB}}}.
\newblock In: {\slshape \bibinfo{booktitle}{Proceedings FMICS}},
  \bibinfo{series}{LNCS 12863}, pp. \bibinfo{pages}{193--201},
  \doi{10.1007/978-3-030-85248-1\_12}.

\bibitemdeclare{article}{DBLP:journals/corr/abs-2108-07878}
\bibitem{DBLP:journals/corr/abs-2108-07878}
\bibinfo{author}{Christiano \surnamestart Braga\surnameend} \&
  \bibinfo{author}{Narciso \surnamestart Mart{\'{\i}}{-}Oliet\surnameend}
  (\bibinfo{year}{2021}): \emph{\bibinfo{title}{B Maude: {A} formal executable
  environment for Abstract Machine Notation Descriptions}}.
\newblock {\slshape \bibinfo{journal}{CoRR}} \bibinfo{volume}{abs/2108.07878}.
\newblock \eprint{2108.07878}.

\bibitemdeclare{inproceedings}{DBLP:conf/birthday/ButlerM13}
\bibitem{DBLP:conf/birthday/ButlerM13}
\bibinfo{author}{Michael~J. \surnamestart Butler\surnameend} \&
  \bibinfo{author}{Issam \surnamestart Maamria\surnameend}
  (\bibinfo{year}{2013}): \emph{\bibinfo{title}{Practical Theory Extension in
  {Event-B}}}.
\newblock In: {\slshape \bibinfo{booktitle}{Theories of Programming and Formal
  Methods - Essays Dedicated to Jifeng He on the Occasion of His 70th
  Birthday}}, pp. \bibinfo{pages}{67--81}, \doi{10.1007/978-3-642-39698-4\_5}.

\bibitemdeclare{inproceedings}{CumborStoddartDunne2010}
\bibitem{CumborStoddartDunne2010}
\bibinfo{author}{David \surnamestart Cumbor\surnameend}, \bibinfo{author}{Bill
  \surnamestart Stoddart\surnameend} \& \bibinfo{author}{Steve \surnamestart
  Dunne\surnameend} (\bibinfo{year}{2010}): \emph{\bibinfo{title}{{Teaching
  Logic Proofs for Formal Aspects}}}.
\newblock In \bibinfo{editor}{Christian \surnamestart Attiogb{\'e}\surnameend}
  \& \bibinfo{editor}{Dominique \surnamestart M{\'e}ry\surnameend}, editors:
  {\slshape \bibinfo{booktitle}{{Colloque "From Research to Teaching Formal
  Methods: The B Method" (TFM-B'2010)}}}, \bibinfo{publisher}{APCB},
  \bibinfo{address}{Nantes, France}, pp. \bibinfo{pages}{2--16}.
\newblock \urlprefix\url{https://hal.science/hal-04993432v1}.

\bibitemdeclare{article}{DBLP:journals/logcom/Fitting98}
\bibitem{DBLP:journals/logcom/Fitting98}
\bibinfo{author}{Melvin \surnamestart Fitting\surnameend}
  (\bibinfo{year}{1998}): \emph{\bibinfo{title}{{leanTAP} Revisited}}.
\newblock {\slshape \bibinfo{journal}{J. Log. Comput.}}
  \bibinfo{volume}{8}(\bibinfo{number}{1}), pp. \bibinfo{pages}{33--47},
  \doi{10.1093/logcom/8.1.33}.

\bibitemdeclare{techreport}{gansner2009drawing}
\bibitem{gansner2009drawing}
\bibinfo{author}{Emden~R. \surnamestart Gansner\surnameend}
  (\bibinfo{year}{2011}): \emph{\bibinfo{title}{Drawing graphs with Graphviz}}.
\newblock \bibinfo{type}{Technical Report}.

\bibitemdeclare{article}{gentzen1935untersuchungen}
\bibitem{gentzen1935untersuchungen}
\bibinfo{author}{Gerhard \surnamestart Gentzen\surnameend}
  (\bibinfo{year}{1935}): \emph{\bibinfo{title}{{Untersuchungen {\"u}ber das
  logische Schlie{\ss}en. I}}}.
\newblock {\slshape \bibinfo{journal}{Mathematische Zeitschrift}}
  \bibinfo{volume}{39}(\bibinfo{number}{1}), pp. \bibinfo{pages}{176--210},
  \doi{10.1007/BF01201353}.

\bibitemdeclare{inproceedings}{DBLP:conf/zum/Grieu24}
\bibitem{DBLP:conf/zum/Grieu24}
\bibinfo{author}{Anne \surnamestart Grieu\surnameend} (\bibinfo{year}{2024}):
  \emph{\bibinfo{title}{From {Event-B} to {Lambdapi}}}.
\newblock In: {\slshape \bibinfo{booktitle}{Proceedings ABZ}}, pp.
  \bibinfo{pages}{387--391}, \doi{10.1007/978-3-031-63790-2\_29}.

\bibitemdeclare{inproceedings}{grieu2025translating}
\bibitem{grieu2025translating}
\bibinfo{author}{Anne \surnamestart Grieu\surnameend},
  \bibinfo{author}{Jean-Paul \surnamestart Bodeveix\surnameend} \&
  \bibinfo{author}{Mamoun \surnamestart Filali\surnameend}
  (\bibinfo{year}{2025}): \emph{\bibinfo{title}{{Translating {Event-B} Models
  and Development Proofs to {TLA+}}}}.
\newblock In: {\slshape \bibinfo{booktitle}{Proceedings ABZ}},
  \bibinfo{series}{LNCS 15728}, \bibinfo{organization}{Springer}, pp.
  \bibinfo{pages}{124--142}, \doi{10.1007/978-3-031-94533-5\_8}.

\bibitemdeclare{inproceedings}{rodin_2025_interactive_trace_replay}
\bibitem{rodin_2025_interactive_trace_replay}
\bibinfo{author}{Jan \surnamestart Gruteser\surnameend} \&
  \bibinfo{author}{Michael \surnamestart Leuschel\surnameend}
  (\bibinfo{year}{2025}): \emph{\bibinfo{title}{{Interactive Trace Replay for
  Event-B Models}}}.
\newblock In: {\slshape \bibinfo{booktitle}{12th Rodin User and Developer
  Workshop}}.
\newblock \urlprefix\url{https://eprints.soton.ac.uk/id/eprint/503334}.

\bibitemdeclare{inproceedings}{xtl_iclp}
\bibitem{xtl_iclp}
\bibinfo{author}{Jan \surnamestart Gruteser\surnameend},
  \bibinfo{author}{Michael \surnamestart Leuschel\surnameend},
  \bibinfo{author}{Katharina \surnamestart Engels\surnameend} \&
  \bibinfo{author}{Fabian \surnamestart Vu\surnameend} (\bibinfo{year}{2026}):
  \emph{\bibinfo{title}{Animation, Verification and Visualisation of Prolog
  Transition Systems with ProB}}.
\newblock In: {\slshape \bibinfo{booktitle}{Proceedings ICLP}},
  \bibinfo{series}{EPTCS}.
\newblock \bibinfo{note}{To appear}.

\bibitemdeclare{article}{hendriks2010teaching}
\bibitem{hendriks2010teaching}
\bibinfo{author}{Maxim \surnamestart Hendriks\surnameend},
  \bibinfo{author}{Cezary \surnamestart Kaliszyk\surnameend},
  \bibinfo{author}{Femke \surnamestart van Raamsdonk\surnameend} \&
  \bibinfo{author}{Freek \surnamestart Wiedijk\surnameend}
  (\bibinfo{year}{2010}): \emph{\bibinfo{title}{Teaching logic using a
  state-of-the-art proof assistant}}.
\newblock {\slshape \bibinfo{journal}{Acta Didactica Napocensia}}
  \bibinfo{volume}{3}.

\bibitemdeclare{article}{iannone2025feels}
\bibitem{iannone2025feels}
\bibinfo{author}{Paola \surnamestart Iannone\surnameend} \&
  \bibinfo{author}{Athina \surnamestart Thoma\surnameend}
  (\bibinfo{year}{2025}): \emph{\bibinfo{title}{`It Feels Like Sort of Cheating
  in Some Ways that You Don’t Fully Show that You’ve Understood the
  Proofs': Mathematics Students Coding Proofs with An Interactive Theorem
  Prover}}.
\newblock {\slshape \bibinfo{journal}{Digital Experiences in Mathematics
  Education}}, pp. \bibinfo{pages}{1--27}, \doi{10.1007/s40751-025-00193-w}.

\bibitemdeclare{inproceedings}{DBLP:conf/asm/IliasovSAR16}
\bibitem{DBLP:conf/asm/IliasovSAR16}
\bibinfo{author}{Alexei \surnamestart Iliasov\surnameend},
  \bibinfo{author}{Paulius \surnamestart Stankaitis\surnameend},
  \bibinfo{author}{David \surnamestart Adjepon{-}Yamoah\surnameend} \&
  \bibinfo{author}{Alexander~B. \surnamestart Romanovsky\surnameend}
  (\bibinfo{year}{2016}): \emph{\bibinfo{title}{Rodin Platform Why3 Plug-In}}.
\newblock In: {\slshape \bibinfo{booktitle}{Proceedings {ABZ} 2016}},
  \bibinfo{series}{LNCS 9675}, pp. \bibinfo{pages}{275--281},
  \doi{10.1007/978-3-319-33600-8\_21}.

\bibitemdeclare{inproceedings}{krings2015failure}
\bibitem{krings2015failure}
\bibinfo{author}{Sebastian \surnamestart Krings\surnameend},
  \bibinfo{author}{Jens \surnamestart Bendisposto\surnameend} \&
  \bibinfo{author}{Michael \surnamestart Leuschel\surnameend}
  (\bibinfo{year}{2015}): \emph{\bibinfo{title}{{From Failure to Proof: The
  ProB Disprover for B and Event-B}}}.
\newblock In: {\slshape \bibinfo{booktitle}{Proceedings SEFM 2015}},
  \bibinfo{series}{LNCS 9276}, \bibinfo{organization}{Springer}, pp.
  \bibinfo{pages}{199--214}, \doi{10.1007/978-3-319-22969-0\_15}.

\bibitemdeclare{inproceedings}{Le08_233}
\bibitem{Le08_233}
\bibinfo{author}{Michael \surnamestart Leuschel\surnameend}
  (\bibinfo{year}{2008}): \emph{\bibinfo{title}{{Declarative Programming for
  Verification: Lessons and Outlook}}}.
\newblock In: {\slshape \bibinfo{booktitle}{Proceedings PPDP'2008}},
  \bibinfo{publisher}{ACM Press}, pp. \bibinfo{pages}{1--7},
  \doi{10.1145/1389449.1389450}.

\bibitemdeclare{inproceedings}{leuschel2020fast}
\bibitem{leuschel2020fast}
\bibinfo{author}{Michael \surnamestart Leuschel\surnameend}
  (\bibinfo{year}{2020}): \emph{\bibinfo{title}{{Fast and Effective
  Well-Definedness Checking}}}.
\newblock In: {\slshape \bibinfo{booktitle}{Proceedings iFM 2020}}, {\slshape
  \bibinfo{series}{LNCS}} \bibinfo{volume}{12546},
  \bibinfo{organization}{Springer}, pp. \bibinfo{pages}{63--81},
  \doi{10.1007/978-3-030-63461-2\_4}.

\bibitemdeclare{inproceedings}{DBLP:journals/corr/abs-2008-02933}
\bibitem{DBLP:journals/corr/abs-2008-02933}
\bibinfo{author}{Michael \surnamestart Leuschel\surnameend}
  (\bibinfo{year}{2020}): \emph{\bibinfo{title}{Prolog for Verification,
  Analysis and Transformation Tools}}.
\newblock In: {\slshape \bibinfo{booktitle}{Proceedings VPT 2020}},
  \bibinfo{series}{{EPTCS} 320}, pp. \bibinfo{pages}{80--94},
  \doi{10.4204/EPTCS.320.6}.

\bibitemdeclare{incollection}{leuschel2023prob}
\bibitem{leuschel2023prob}
\bibinfo{author}{Michael \surnamestart Leuschel\surnameend}
  (\bibinfo{year}{2023}): \emph{\bibinfo{title}{{ProB: Harnessing the Power of
  Prolog to Bring Formal Models and Mathematics to Life}}}.
\newblock In: {\slshape \bibinfo{booktitle}{Prolog: The Next 50 Years}},
  \bibinfo{series}{LNCS 13900}, \bibinfo{publisher}{Springer}, pp.
  \bibinfo{pages}{239--247}, \doi{10.1007/978-3-031-35254-6\_19}.

\bibitemdeclare{article}{DBLP:journals/sttt/LeuschelB08}
\bibitem{DBLP:journals/sttt/LeuschelB08}
\bibinfo{author}{Michael \surnamestart Leuschel\surnameend} \&
  \bibinfo{author}{Michael~J. \surnamestart Butler\surnameend}
  (\bibinfo{year}{2008}): \emph{\bibinfo{title}{{ProB}: an automated analysis
  toolset for the {B} method}}.
\newblock {\slshape \bibinfo{journal}{STTT}}
  \bibinfo{volume}{10}(\bibinfo{number}{2}), pp. \bibinfo{pages}{185--203}.
\newblock \urlprefix\url{http://dx.doi.org/10.1007/s10009-007-0063-9}.

\bibitemdeclare{article}{lincoln1989adventures}
\bibitem{lincoln1989adventures}
\bibinfo{author}{Patrick \surnamestart Lincoln\surnameend} \&
  \bibinfo{author}{Jim \surnamestart Christian\surnameend}
  (\bibinfo{year}{1989}): \emph{\bibinfo{title}{{Adventures in
  Associative-Commutative Unification}}}.
\newblock {\slshape \bibinfo{journal}{Journal of Symbolic Computation}}
  \bibinfo{volume}{8}(\bibinfo{number}{1-2}), pp. \bibinfo{pages}{217--240},
  \doi{10.1016/S0747-7171(89)80026-4}.

\bibitemdeclare{inproceedings}{DBLP:conf/nfm/OlarteORR25}
\bibitem{DBLP:conf/nfm/OlarteORR25}
\bibinfo{author}{Carlos \surnamestart Olarte\surnameend},
  \bibinfo{author}{Daniel \surnamestart Osorio\surnameend},
  \bibinfo{author}{Carlos \surnamestart Ram{\'{\i}}rez\surnameend} \&
  \bibinfo{author}{Camilo \surnamestart Rocha\surnameend}
  (\bibinfo{year}{2025}): \emph{\bibinfo{title}{Algorithmic Analysis of
  {Event-B} in Rewriting Logic}}.
\newblock In: {\slshape \bibinfo{booktitle}{Proceedings {NFM} 2025}},
  \bibinfo{series}{LNCS 15682}, pp. \bibinfo{pages}{275--293},
  \doi{10.1007/978-3-031-93706-4\_16}.

\bibitemdeclare{article}{DBLP:journals/aicom/RiazanovV02}
\bibitem{DBLP:journals/aicom/RiazanovV02}
\bibinfo{author}{Alexandre \surnamestart Riazanov\surnameend} \&
  \bibinfo{author}{Andrei \surnamestart Voronkov\surnameend}
  (\bibinfo{year}{2002}): \emph{\bibinfo{title}{The design and implementation
  of {VAMPIRE}}}.
\newblock {\slshape \bibinfo{journal}{{AI} Commun.}}
  \bibinfo{volume}{15}(\bibinfo{number}{2-3}), pp. \bibinfo{pages}{91--110}.
\newblock
  \urlprefix\url{http://content.iospress.com/articles/ai-communications/aic259}.

\bibitemdeclare{phdthesis}{DBLP:phd/basesearch/Schmalz12}
\bibitem{DBLP:phd/basesearch/Schmalz12}
\bibinfo{author}{Matthias \surnamestart Schmalz\surnameend}
  (\bibinfo{year}{2012}): \emph{\bibinfo{title}{Formalizing the logic of
  event-B: Partial functions, definitional extensions, and automated theorem
  proving}}.
\newblock Ph.D. thesis, \bibinfo{school}{{ETH} Zurich, Z{\"{u}}rich,
  Switzerland}, \doi{10.3929/ETHZ-A-007577749}.
\newblock \urlprefix\url{https://hdl.handle.net/20.500.11850/64337}.

\bibitemdeclare{unpublished}{stolze:hal-04398119}
\bibitem{stolze:hal-04398119}
\bibinfo{author}{Claude \surnamestart Stolze\surnameend},
  \bibinfo{author}{Olivier \surnamestart Hermant\surnameend} \&
  \bibinfo{author}{Romain \surnamestart Guillaum{\'e}\surnameend}
  (\bibinfo{year}{2024}): \emph{\bibinfo{title}{{Towards Formalization and
  Sharing of Atelier B Proofs with Dedukti}}}.
\newblock \urlprefix\url{https://hal.science/hal-04398119}.
\newblock \bibinfo{note}{Working paper or preprint}.

\bibitemdeclare{inproceedings}{visb}
\bibitem{visb}
\bibinfo{author}{Michelle \surnamestart Werth\surnameend} \&
  \bibinfo{author}{Michael \surnamestart Leuschel\surnameend}
  (\bibinfo{year}{2020}): \emph{\bibinfo{title}{{VisB: A Lightweight Tool to
  Visualize Formal Models with SVG Graphics}}}.
\newblock In: {\slshape \bibinfo{booktitle}{Proceedings ABZ}}, {\slshape
  \bibinfo{series}{LNCS}} \bibinfo{volume}{12071},
  \bibinfo{publisher}{Springer}, pp. \bibinfo{pages}{260--265},
  \doi{10.1007/978-3-030-48077-6\_21}.

\end{thebibliography}

\appendix

\noindent The source code is available in \texttt{src/sequent\_prover} as part of ProB’s source code: \url{https://stups.hhu-hosting.de/downloads/prob/source}.

\end{document}